\documentclass[reprint,amssymb,amsmath,aps,prl,nofootinbib,longbibliography]{revtex4-1}

\usepackage{graphicx}
\usepackage{dcolumn}
\usepackage{bm}
\usepackage{hyperref}
\usepackage{amsthm}
\usepackage{relsize}
\usepackage{natbib}

\usepackage[abs]{overpic}

\usepackage{xcolor,varwidth}

\newcommand{\h}{\hbar}

\newcommand{\kB}{k_B}

\newcommand{\be}{\begin{equation}}
\newcommand{\ee}{\end{equation}}

\begin{document}


\title{
Large quantum superpositions of a nanoparticle immersed in superfluid helium}

\author{O. Lychkovskiy}
\affiliation{ Russian Quantum Center, Novaya St. 100A, Skolkovo, Moscow Region, 143025, Russia.}


\begin{abstract}
Preparing and detecting spatially extended quantum superpositions of a massive object comprises an important fundamental test of quantum theory. These quantum states are extremely fragile and tend to quickly decay into incoherent mixtures due to the environmental decoherence. Experimental setups considered up to date address this threat in a conceptually straightforward way --  by eliminating the environment, i.e. by isolating an object in a sufficiently high vacuum. We show that another option exists: decoherence is suppressed in the presence of a strongly interacting environment if this environment is superfluid. Indeed, as long as an object immersed in a pure superfluid at zero temperature moves with a velocity below the critical one, it does not create, absorb or scatter any excitations of the superfluid. Hence, in this idealized situations the decoherence is absent. In reality the decoherence will be present due to thermal excitations of the superfluid and impurities contaminating the superfluid. We examine various decoherence channels in the superfluid $^4$He.  It is shown that the total decoherence time can be as large as tens of seconds for a $10^6$ amu nanoparticle delocalized over $300$ nm in helium at $1$ mK. Performing interference experiments in superfluid helium can provide certain practical advantages compared to conventional schemes, e.g. compensation of gravity by the buoyancy force and effective cooling.
\end{abstract}


\maketitle




There is an ongoing activity in preparing and detecting spatially extended coherent quantum states of a massive nanoscale object \cite{hornberger2012colloquium,arndt2014testing}. The goal of this activity is to test superposition principle which lies in the heart of quantum theory but 
apparently clashes with the general theory of relativity  \cite{penrose1996gravity}.
Interference of nanometer-sized molecules with the mass on the order of $10^4$ amu on an optical grating with the period of 266 nm has been demonstrated in recent experiments \cite{gerlich2011quantum, eibenberger2013matter}. Experimental setups aimed at pushing the mass of interfering nanoparticles by several orders of magnitude  are being discussed and developed \cite{hornberger2012colloquium,arndt2014testing,nimmrichter2011concept,romero2011large,kaltenbaek2012macroscopic,kaltenbaek2015macroscopic,bateman2014near,dorre2014photofragmentation}. Alternative ways of probing nonlocal quantum states of levitating nanoparticles based on entangling center-of-mass motion with another (external or internal) degree of freedom  have also been proposed \cite{romero2010toward,chang2010cavity,romero2011optically,romero2012quantum,scala2013matter,yin2013large}

It is well understood that the major (and maybe the only) threat for nonclassical superpositions is environmental decoherence  \cite{joos1985emergence}. In the context of the state-of-the-art interference experiments with nanoparticles decoherence is mainly due to collisions with ambient gas molecules \cite{arndt2014testing}. In experimental schemes implemented or proposed up to date this threat is addressed in a straightforward way: To suppress environmental decoherence one eliminates the environment, i.e. isolates an object in a sufficiently high vacuum.  For example,  in the record-breaking experiments \cite{gerlich2011quantum, eibenberger2013matter} the pressure was less than $10^{-8}$ mbar. Increasing the mass (and, hence, the size) of the nanoparticles will require progressively better vacuum \cite{nimmrichter2011testing}.

In the present paper we argue that vacuum is not the only low-decoherence host medium: Superfluid helium is an alternative. Indeed, as long as an object immersed in a pure superfluid at zero temperature moves with a velocity less than the critical one, it does not create, absorb or scatter any excitations of the superfluid. Consequently, in this idealized situations the decoherence is completely absent, despite the fact that the object is surrounded by the dense medium and strongly interacts with it. Simply put, in this situation one eliminates decoherence without eliminating the environment.

In a realistic situation a spatially extended superposition of a nanoparticle immersed in superfluid $^4$He at a finite temperature will decohere with a certain rate. The reasons for the decoherence include interactions with thermal excitations of the superfluid and scattering of the $^3$He atoms which inevitably contaminate the superfluid  $^4$He. We examine various sources of decoherence and estimate corresponding decoherence times.
These estimates show that the total decoherence rate can be low enough to sustain large quantum superpositions for times sufficient to reveal these superpositions experimentally.

To make quantitative estimates we choose specific reference values for physical quantities involved. The mass of the nanoparticle is taken to be $M\sim 10^6$ amu, which corresponds to  sizes $a\sim (5-10)$ nm, depending on the density $\rho$ of the nanoparticle.  The coherence length (i.e. the ``size'') of the superposition is taken to be $D \sim 300$ nm. These figures are motivated by state-of-the-art experiments and recent proposals \cite{arndt2014testing,romero2011large,bateman2014near}. Unless explicitly specified, the nanoparticle is assumed to be a homogeneous sphere, $a$ being its radius. Effects of non-sphericity will be considered only when discussing rotation of the nanoparticle.
Temperature of the superfluid helium is considered to be $T\sim 1$ mK.

Now we are in a position to estimate contributions of various sources of decoherence. The estimates are essentially based on theory summarized in the book \cite{schlosshauer2008decoherence}. We note, however, that this theory should be applied not to the original Hamiltonian of the system but to the low-energy effective Hamiltonian which describes the superfluid helium as a collection of quasiparticles and treats the nanoparticle in a polaron-like manner \cite{KhalatnikovBook}.


{\em Scattering of} $^3$He {\em  impurities.}
Natural helium contains a $\sim 10^{-6}$ fraction of $^3$He. A method to purify $^4$He up to a relative concentration $X_3\equiv n_3/(n_4+n_3) \leq 0.5 \cdot 10^{-15} $ (where $n_4$ and $n_3$ are number densities of $^4$He and $^3$He, respectively)   is well established \cite{mcclintock1978apparatus}. Remarkably, the latter figure reflects the lack of technique to measure small concentration of $^3$He rather than the ultimate concentration \cite{mcclintock1978apparatus}. We choose the reference concentration of $^3$He impurities to be $10^{-15}$.

At low temperatures and concentrations $^3$He impurities can be considered as a dilute ideal gas. Effective wavelength of $^3$He impurities reads
\be
\lambda_3=\frac{2\pi \h}{ \sqrt{3 m_3^{\rm eff} \kB T}}\simeq 30\,{\rm nm}
\ee
with $m_3^{\rm eff}\simeq 2.34 m_3$ \cite{anderson1966thermal,bardeen1967effective}, $m_3$ being the mass of the $^3$He atom.
Since $a \ll \lambda_3 \ll D$, one can use the short wavelength limit described in Ref. \cite{schlosshauer2008decoherence} along with the diffractive cross section to calculate the decoherence time, with the result
\begin{align}\label{decoherence time He3}
\tau_{{\rm He}3}
\simeq &
\frac{  m_3^{\rm eff} }{8 \,a^2\, n_3\sqrt{2\, \pi\, m_3^{\rm eff} \kB T}}
 \simeq 40 \, {\rm s} \, \times
 \\
 \left(\frac{ M}{10^6 \, {\rm amu}}\right)^{-2/3} \nonumber
 &   \left(\frac{\rho}{1 {\rm g}/ {\rm cm}^3 }\right)^{2/3}  \left(\frac{T}{1 \, {\rm mK}}\right)^{-1/2}\, \left(\frac{X_3}{10^{-15} }\right)^{-1}.
\end{align}
As will be seen in what follows, scattering on the $^3$He impurities is the dominant source of decoherence for the reference values of parameters.


{\em Non-resonant scattering of thermal phonons.} Effective wavelength of phonons reads
\be
\frac{2\pi \h v_s}{8 \kB T  }\simeq 1400~{\rm nm},
\ee
where $v_s \simeq 238\,{\rm m}/{\rm s}$ is the sound velocity in the superfluid helium and factor 8 in the denominator appears from a certain integration over the phase space \cite{schlosshauer2008decoherence}. This wavelength is much larger than $a$ and $D$, hence the decoherence occurs in the long-wavelength regime \cite{schlosshauer2008decoherence}. The decoherence time reads
\begin{align}\label{decoherence time phonons}
&\tau_{ph}=
 \frac{54\pi}{11\cdot 8! \, \zeta(9)} \hbar^{9} v_s^{8} D^{-2} a^{-6} (\kB T)^{-9}\simeq 2.5\cdot 10^{4}\, {\rm s} \,\times\\ \nonumber
 &\left(\frac{M}{10^6 \, {\rm amu}}\right)^{-2}  \left(\frac{\rho}{1 {\rm g}/ {\rm cm}^3 }\right)^2 \left(\frac{D}{300 \, {\rm nm}}\right)^{-2} \left(\frac{T}{1 \, {\rm mK}}\right)^{-9}.
\end{align}
This expression is obtained by considering the nanoparticle as a rigid sphere with the  phonon scattering cross section borrowed from \cite{landau2003fluid}. Accounting for elastic properties of the nanoparticle results in a prefactor which is generically on the order of one.

Observe a strong dependence of the decoherence time on temperature and mass of the nanoparticle. In particular, $\tau_{ph}$ becomes an order of magnitude smaller than $\tau_{{\rm He}3}$ already at $T=3$ mK (and reference values for other parameters).

Two sources of decoherence considered above are the only ones of practical relevance. However, for the sake of completeness we briefly discuss other sources of decoherence below.

{\em Frozen modes of nanoparticle and superfluid}.  Absorption and radiation of phonons in resonance with vibrational modes of the nanoparticle are not taken into account in eq. \eqref{decoherence time phonons} and, generally speaking, should be considered separately. However vibrational modes of the nanoparticle are completely frozen out at our reference temperature of 1 mK. Indeed, the lowest vibrational eigenenergy is on the order of
$
\hbar c/a
$
, $c$ being the speed of sound of material the nanoparticle is made from. This corresponds to energy $\sim 1$ K for $c \sim 10^3$ m$/$s.

Elementary excitations of the superfluid helium other than phonons, namely rotons and vortex rings, also have typical energies of $\sim 1$ K \cite{donnelly1991quantized} and thus can be disregarded.

{\em  Nanoparticle  rotation.} Rotation of the nanoparticle is not frozen out at 1 mK, in contrast to vibration. Indeed, thermal average of the angular momentum $L$ of the nanoparticle is rather large,
\begin{align}
L= &\sqrt{3 I \kB T}\simeq\\
&  365 \hbar \left(\frac{M}{10^6 \, {\rm amu}}\right)^{5/6} \left(\frac{\rho}{1 {\rm g}/ {\rm cm}^3 }\right)^{-1/3} \left(\frac{T}{1 \, {\rm mK}}\right)^{1/2}, \end{align}
where
$I$
is the moment of inertia of the nanoparticle. Transitions between rotational levels of the nanoparticle are accompanied by resonant absorption and emission of phonons which contribute to decoherence. To estimate this contribution we assume that the nanoparticle is not an ideal sphere but rather an ellipsoid with a small ellipticity parameter $\varepsilon$. This simple assumption will suffice to reveal relevant physics. The semi-axes of the ellipsoid are equal to $a$ up to relative corrections $\sim \varepsilon$.

Wavelength of a resonantly emitted or absorbed phonon reads
\be\label{wavelengh rotation}
\lambda_{res}=\frac{2\pi v_s I}{ L}\simeq 1.4\cdot 10^6 \,{\rm nm}.
\ee
One can see that the decoherence occurs in the long wavelength regime. Hence the decoherence time is $\sim(\lambda_{res}/D)^2$ times larger than the life time of an excited rotational state. The latter can be estimated following Ref. \cite{landau2003fluid}.  The resulting estimate for the decoherence time reads
\begin{align}\label{decoherence time rotation}
&\tau_{rot}\sim \left(\frac{\lambda_{res}}{D}\right)^2 \frac{\hbar v_s^3 I^5}{\rho_{\rm He} \varepsilon^2 a^8 L^5} \sim 10^{13}\, {\rm s}\, \times \, \varepsilon^{-2} \, \times \\ \nonumber
 &  \, \left(\frac{M}{10^6 \, {\rm amu}}\right)^{19/6}  \left(\frac{\rho}{1 {\rm g}/ {\rm cm}^3 }\right)^{1/3} \left(\frac{D}{300 \, {\rm nm}}\right)^{-2} \left(\frac{T}{1 \, {\rm mK}}\right)^{-7/2},
\end{align}
where $\rho_{\rm He}$ is the density of helium.
Quite remarkably, despite large angular momentum possessed by the nanoparticle, rotation contributes negligibly to the decoherence. This is because resonantly emitted and absorbed phonons have an extremely long wavelength, see eq. \eqref{wavelengh rotation},  and thus probe the position of the nanoparticle with an extremely low resolution.



{\em Discussion.} Robustness of large spatial superpositions of an object surrounded by and strongly interacting with superfluid helium -- a medium with number density
$2\cdot10^{22}$ atoms per cm$^3$
--  is quite unusual. Indeed, if superfluid helium would be substituted by a noninteracting Bose gas with the same density and mass of an individual boson (in other words, if interactions between helium atoms would be ``switched of''), the decoherence time would drop from $40$ s (see eq. \eqref{decoherence time He3}) to less than $10^{-10}$ s (both figures are given for reference values of parameters)! The aforementioned remarkable robustness is a new facet of a far-reaching similarity between superfluid and vacuum \cite{volovik2009universe}.

Revealing a spatial superposition in interference experiment requires that the superposition is sustained for a certain time. E.g. in a typical Talbot-Law setup this is a  Talbot time \cite{hornberger2012colloquium}
\be\label{Talbot time}
\tau_T=\frac{M D^2}{2\pi \hbar} \simeq 0.2 \, {\rm s} \,\times \left(\frac{M}{10^6 \, {\rm amu}}\right) \left(\frac{D}{300 \, {\rm nm}}\right)^{2} .
\ee
One can see that decoherence times due to relevant sources of decoherence given by eqs. \eqref{decoherence time He3} and \eqref{decoherence time phonons} are well above the Talbot time for the reference values of parameters. Moreover, enough space is left for various trade-offs which can result in increasing the mass of the nanoparticle well above $10^6$ amu, increasing the coherence length of the superposition  or relaxing experimental requirements on temperature and concentration of $^3$He impurities.

An additional requirement is that the nanoparticle should not reach a velocity exceeding the critical velocity in superfluid helium during the experimental run. This requirement is satisfied for a nanoparticle gravitationally accelerated for times suggested by eq. \eqref{Talbot time}.


We have shown that an interference experiment with a nanoparticle immersed in superfluid helium is in principle feasible. Clearly, performing such an experiment can be technically challenging. However, there can be certain pay-offs compared to conventional schemes relying on high vacuum. First, cooling a nanoparticle to mK temperatures (which is essential for some schemes \cite{bateman2014near}) requires rather sophisticated techniques in vacuum but is seamless in superfluid helium. Second, compensation of gravity force is often desirable in interference experiments with massive nanoparticles in order to keep the free fall distance of the nanoparticle within the apparatus dimensions \cite{nimmrichter2011concept,arndt2014testing} (in particular, this is one of the motivations of  the space-based proposal MAQRO \cite{kaltenbaek2012macroscopic,kaltenbaek2015macroscopic}).  If one is able to manufacture nanoparticles with density $\rho$ approximately equal to the density of liquid helium, $ \rho_{\rm He} \simeq 0.145$ g$/$cm$^3$, it becomes possible to approximately compensate the gravity force by the buoyancy force.

A brief remark on a potential metrological application of interference of nanoparticles immersed in superfluid $^4$He is in order. Since scattering on $^3$He impurities is the dominant source of decoherence for temperatures below few mK, dependence of the decoherence time on the concentration of  $^3$He, see eq. \eqref{decoherence time He3}, can be used to measure this concentration beyond the limitations imposed by current techniques  \cite{mcclintock1978apparatus,hendry1987continuous}.

{\em Summary.} To summarize, we have studied decoherence of a spatially extended quantum state of a massive ($M \gtrsim 10^6$ amu) nanoparticle immersed in superfluid helium at $T\sim 1$ mK.  We have shown that a coherent delocalization on the order of $\sim 300$ nm can be sustained for tens of seconds, which is more than enough to reveal the nonclassical nature of the nanoparticle state in an interference experiment.

Two sources of decoherence have been found to be of practical relevance -- scattering of $^3$He impurities and thermal phonons.  Accordingly, decreasing concentration of $^3$He impurities and temperature of the superfluid helium are primary measures to probe even higher nanoparticle masses and larger delocalization.

\begin{acknowledgments}
{\em Acknowledgements.}
The author is grateful to M. Aspelmeyer, J. Cotter, A. Fedorov, M. Kagan, P. McClintock, G. Pickett, G. Shlyapnikov, V. Tsepelin, H. Ulbricht  and M. Zvonarev for fruitful discussions and valuable remarks. The work has been supported by the RFBR under the grant N$^{\rm o}$ 16-32-00669.
\end{acknowledgments}

\bibliography{D:/Work/QM/Bibs/macrosuperpositions,D:/Work/QM/Bibs/decoherence,D:/Work/QM/Bibs/polaron,D:/Work/QM/Bibs/He,D:/Work/QM/Bibs/phonon_scattering}

\end{document}